\documentstyle[11pt,aaspp4]{article}
\received{}
\revised{}
\accepted{}
\cpright{}{}
\journalid{}{}
\articleid{}{}
\paperid{}
\cpright{}{}
\ccc{}
\lefthead{J. Jurcsik}
\righthead{Spatial metallicity asymmetry in  $\omega$ Centauri}
\begin{document}
\title{Spatial metallicity asymmetry in $\omega$ Centauri}
\author{Johanna Jurcsik}
\affil{Konkoly Observatory, P.O.~Box~67, H--1525,
Budapest, Hungary \\
E-mail: jurcsik@buda.konkoly.hu}
\authoraddr{Konkoly Observatory, P.O.~Box~67, H--1525,
Budapest, Hungary}
%
%
\begin{abstract}
The compilation of the metallicity values of $\omega$~Centauri stars led to  
detection of spatial asymmetry in the metallicity distribution of the bright 
giants. It was found that the most and least metal-poor objects are separated 
along galactic latitude. Statistical tests show that the chance occurrence 
of the phenomenon is very low. The subgiant and the RR~Lyrae samples do not show, 
however, similar segregation of the different metallicity objects. The physical 
parameters of the variables calculated by using empirical formulae indicate that 
the bulk of the RRab stars comprises a very homogeneous group regarding both 
their metallicity and mass values. These stars are most probably already in the 
evolved stage of their horizontal branch evolution. 
\end{abstract} 
%
%
\section{Introduction}
One of the most peculiar globular clusters is $\omega$~Centauri with its 
extreme characteristics. It is not only the largest, most massive, and most 
oblate Galactic globular cluster but it is also one of the least bounded 
clusters with $\sim 30$ Gyr half-mass relaxation time (Meylan et~al. 1995). 
Accordingly, the dynamical evolution of the cluster is extremely slow. The 
large metallicity spread observed (greater than 1.5~dex) is also unique in 
globular clusters. Regarding some of its characteristics $\omega$~Cen resembles 
the dwarf spheroidal galaxies rather than the globular clusters. The large 
number of the variables also gives useful information on the stellar populations 
of the system.

The anomalous properties and the relatively short distance of the cluster
stimulated many extensive photometric and spectroscopic studies. In spite of these 
efforts, the chemical and dynamical history of the system are not yet understood 
clearly. There is evidence indicating that both primordial and evolutionary 
effects are responsible for the origin of the metallicity spread and for the other 
observed peculiarities of the chemical compositions (see e.g. Norris \& Da Costa 
1995, Suntzeff \& Kraft 1996). Different models involving two periods of star 
formation, self-enrichment and also the possibility of two merging components were 
suggested to explain the systematics found in the chemical compositions and in the 
radial distribution of the different groups of stars; some chemical composition 
indicators (Ca~lines, CN~bands) clearly show evidence of radial gradient. Norris 
et~al. (1997) found that the larger metallicity giants form a radially more 
concentrated group than the metal-poor ones, and do not follow the same systemic 
rotation as the metal-poor component exhibits. They concluded that this situation 
favors the merging scenario. 

The large number of the accurate metallicity measurements makes it possible to 
map the unbiased spatial metallicity distribution in $\omega$~Centauri. 
The aim of all such previous investigations were to find differences between 
the radial distributions of the different metallicity stars, thus the true 
metallicity map of the cluster has not yet been actually constructed. In this  
Letter it is shown that there is not only a radial gradient in the metallicity 
distribution (see Norris et~al. 1997), but the most and the least metal-poor 
bright giants tend to be located at the opposite sides of the cluster. 
It is also indicated that the bulk of the RR~Lyrae stars belongs to a very
homogeneous population.
%
%
\section{The data}
All the published metallicity values of $\omega$~Centauri stars have been 
collected. The two extended metallicity surveys (Norris, Freeman \& Mighell 
1996, NFM, and Suntzeff \& Kraft 1996, SK) included most of those stars which 
were analyzed in the other works, thus to minimize the effect of discrepancies 
between the metallicity scales, for those stars belonging to the NFM and/or SK 
samples only these observations are considered. The [Ca/H] values of NFM are 
transformed to [Fe/H] using the high dispersion results of 40 red giants by 
Norris and Da Costa (1995, ND). For these stars the [Fe/H] values given in ND, 
whereas for the others the averages of the NFM and SK observations are used. 
Additional data include the 3 chemically peculiar (Ba, S) stars analyzed by Vanture, 
Wallerstein \& Brown (1994), the three variables and ROA\,24 from Gonzalez \& 
Wallerstein (1994), and the RR Lyrae stars with known metallicity. For 48 singly 
periodic RRab stars the [Fe/H] values are determined from the light curve parameters 
using the OGLE observations (Kaluzny et~al. 1997). This method give $\sim0.15$~dex 
accurate [Fe/H] values (Jurcsik \& Kov\'acs 1996) on a metallicity scale (Jurcsik 
1995) calibrated to high dispersion spectroscopic results consistent with the ND 
metallicity scale. Besides these data, $\Delta S$ observations of 35 additional 
(mostly RRc) variables transformed to the above metallicity scale from the studies 
of Butler, Dickens \& Epps (1978, BDE), and Gratton, Tornambe \& Ortolani 
(1986, GTO) are also used.
 
The above compilation comprises metallicities of altogether 792 stars. 
The majority of the data comes from SK and NFM. Both of these samples aimed to 
give the best possible representation of the unbiased distribution of the stellar
metallicities, therefore only the brightness was used as a selection criterion 
for the bright giants. Concerning our present result, the most important is, however, 
that there was no selection effect indicated in any of the data used which would 
provide a selection bias toward the position of the different metallicity stars 
within the cluster, besides limitations in their radial distances from the cluster 
center. Thus, it can be assumed that the above compilation reflects the true spatial
 distribution of the different metallicity bright giants and there is no clear 
indication of coordinate related sampling in the case of the variables and 
subgiants, too.
%
%
\section{The metallicity map}
Using the metallicity values of the 792 stars referred in Sec.~2 the apparent 
location of the different metallicity stars within the cluster can be plotted. 
We use the ROA catalogue positions (Woolley 1966) and for additional stars the 
coordinates given by NFM and Kaluzny et~al. (1997) all of them transformed into 
galactic coordinate system (J2000). Applying different metallicity criteria and 
magnitude limits it is found that the most and the least metal-poor giants seem 
to be located at different parts of the cluster. The effect is the most pronounced 
for the brightest giants (359 stars with $V \le 12.75$~mag) if stars with [Fe/H] 
$\le -1.75$ and [Fe/H] $\ge -1.25$ are selected. In this case the most metal-poor 
and the most metal-rich stars are located as shown in Fig.~1. The non-uniform 
distribution of the metallicity extremes is rather striking in this figure, the 
most metal-poor sample belongs mostly to the `upper right' part of the cluster 
whereas the metal-rich one to the opposite side. There is no metallicity extreme 
star below or above a given latitude limit in this bright sample. The separation 
between the centers of these groups is 6.2'.

Though the separation of the extreme metallicity stars is clearly evident in 
Fig.~1, it is also necessary to check whether such metallicity distribution could 
also occur just by chance or it is an artifact of the small sample sizes. In order 
to decide the real significance of the phenomenon the following test is performed. 
The coordinates and metallicities of stars brighter than a magnitude limit are 
randomly coupled in sufficiently large number (5000) of simulations. In each 
simulation the same metal-poor and metal-rich samples are selected as in Fig.~1 and 
the distances ($\Delta C$) between their mean coordinates (centers) are determined. 
In Fig.~2 the histogram of the distribution of the $\Delta C$ values obtained from 
this test is shown, when 464 stars brighter than 13~mag are used. In this case 55 
and 37 stars belong to the metal-poor and metal-rich groups, respectively. The 
arrow indicates the observed value of $\Delta C$. We can draw the conclusion from 
this test that such separation between the different metallicity samples as observed 
has very low probability. The measured distance between the samples was reached only 
in less than 0.1\% of the realizations. Performing this test by using other magnitude 
and metallicity criteria, it is found that even if no magnitude selection is applied 
the observed separation is reached only in about 5\% of the simulations. When using 
the bright ($V\le 12.75$~mag) sample the probability is, however, reduced practically 
to zero. While the observed value in this case is 6.2', none of the 5000 simulations 
resulted in $\Delta C$ value larger than 6.0'. Concerning the metallicity ranges 
chosen, if the two different metallicity groups comprise about $50-50$\% of the 
bright giant sample then the observed separation is still statistically significant 
with 7\% probability.

The significance of the difference between the distributions of the extreme metallicity 
groups can also be determined by using two-sample Kolmogorov-Smirnov test. Assuming 
as null hypothesis that their latitude distributions are the same, the K-S test for 
the bright samples ($V\le 12.75$ and 13.00~mag) gives 0.0005 and 0.0020 probability 
to reach the observed distribution if the null hypothesis is true. When the whole 
sample is regarded with no magnitude limit then the null hypothesis is firmly rejected 
with p-value~$=0.0$. The maximum values of the differences between the cumulative 
distribution functions are 0.5, 0.4 and 0.3 in the above cases, respectively.

It is also worth noting that the direction of the separation has two special 
properties. Firstly, the metal richer population is located towards the galactic 
disc, pointing to the galactic center. Secondly, the photometric minor axis of 
$\omega$~Cen, which is shown by dashed line in Fig.~1, is very close to the 
direction of galactic longitude, indicating that the rotational axis of the 
cluster is nearly perpendicular to the separation. Such coincidences further 
reduce the probability of the chance occurrence of the phenomenon.

To see the separation of the metallicity extremes of the different groups of the
stars, in Fig.~3 the relative galactic latitudes $\Delta {\rm b}$ measured from 
the position of the cluster center (${\rm b}=14.98^{\circ}$) are plotted against 
the brightness. As in Fig.~1, filled and open circles indicate the most and least 
metal-poor stars, respectively. This figure also shows that in the bright sample 
there is no metal rich star above $\Delta {\rm b}\simeq 3$', while just the 
opposite is true for the most metal-poor stars. There is no $V < 12.75$~mag 
most metal-poor giant below $\Delta {\rm b}\simeq -4$', and the three stars brighter 
than 13~mag which lie at law latitudes have just the limit value of the metallicity ($-1.75\div-1.77$~dex) as far as the selection criterion is concerned.

The faint sample which comprises most of the variables and the subgiants, does 
not show a similar segregation of the most and least metal-poor stars as observed 
for the bright sample. The extremes are equally distributed at both sides of the 
cluster. This sample, however, also shows a special feature in the metallicity 
distribution. There is no most metal deficient subgiant close to the $\Delta 
{\rm b}=0$ plane, at the position most of them are found in the bright sample. 
Although, the sample of subgiants may suffer from some selection bias (e.g. all 
the subgiants measured are outside a central radius of 8', only stars with 
small proper motion values were observed, etc), we do not see any specific reason 
for such artifact. The [Fe/H] values of the faint sample are, however, less 
certain than those of the brighter stars, thus it is also possible that the 
$\sim 0.2$~dex metallicity range of the most metal-poor stars does not allow 
correct selection of the objects belonging to this sample. Therefore, the reality 
of this anomaly cannot be taken for sure at present. Consequently, avoiding 
misleading interpretations we only call for studies searching very metal-poor 
subdwarfs close to the `equatorial' region of the cluster. It also has to be 
mentioned, however, that K-S test of the faint sample accepts the null hypothesis 
at only 5\% significance level.
%
%
\section{The RR Lyrae stars}
The metallicities of the RR Lyrae stars are calculated from their light curve 
parameters (empirical values), and also come from spectroscopic observations. 
Comparing the empirical and spectroscopic data, however, significant discrepancies 
can be found. There are 23 RRab stars in the BDE sample which also have empirical 
[Fe/H] values. The mean difference between the two types of their metallicities is 
only $-0.08$~dex (${\rm [Fe/H]_{emp}-[Fe/H]_{sp}}$), but with unexpectedly large, 
0.52~dex standard deviation. Comparing the empirical data with GTO's observations 
(10 stars in common) the mean difference is $-0.06$~dex with a bit smaller 0.38~dex 
rms scatter. Therefore, when only spectroscopic results are available but from 
both sources, GTO's data are used. Since the empirical values are assumed to be 
$0.10 - 0.15$~dex correct (see Jurcsik \& Kov\'acs 1996, Layden 1997, Kaluzny 
et~al. 1998), it is suspected that the spectroscopic data are not accurate enough. 
The discrepancies between BDE's and GTO's results also indicate uncertainties
of the $\Delta S$ observations. At present, however, we cannot really decide 
whether the spectroscopic data are so much uncertain or the empirical formula 
calculating the metallicity is not valid for the $\omega$~Centauri variables. 
However, before speculating on any such interpretation, it would be very important 
to obtain high accuracy spectroscopic observations of the variables in order to 
confirm the discrepancy. 

Unfortunately, this uncertainty questions the reality of the metallicity map of 
the variables. Thus we only mention that when using exclusively the empirical 
data there are 1 and 7 stars belonging to the most and least metal-poor groups, 
respectively. The most metal-poor one (V99) lies very close to the center and all 
the least metal-poor ones are between $14.85^\circ$ and $15.04^\circ$ galactic 
latitudes. This distribution does not contradict the separation found in the case 
of giants, but does not significantly confirm it either. If the spectroscopic 
metallicities are also involved, no clear evidence of any separation can be 
found, as it can be seen in Fig.~3 where most of the stars in the $14 \leq V \leq 15$ 
magnitude range are RR~Lyraes. But as the discrepancy between the spectroscopic 
and empirical metallicity values are surprisingly large we do not put too much 
confidence in this result.

The application of the physical parameter -- light curve relations (Kov\'acs \& 
Jurcsik 1996, 1997; Jurcsik 1998), besides [Fe/H], yields luminosity, 
temperature, and mass values of the variables, too. Thus we can construct the 
HR diagram of these stars as shown in Fig.~4. The metallicity extremes are 
denoted by open circles with size proportional to their [Fe/H]. The average 
[Fe/H] of the remaining 40 variables is $-1.54$~dex with 0.08~dex standard 
deviation. These stars (shown as filled circles) seem to be located along a 
similar track as evolutionary model calculations predict. For comparison a 
segment of the oxygen enhanced evolutionary track (${\rm [Fe/H]}=-1.48$; 
$M=0.66M_{\odot}$) of Dorman (1992) is also indicated. The empirical $\log L$ 
and $\log T$ values are shifted by 0.10 and 0.015, respectively, in order to 
reach agreement with evolutionary results (see also Jurcsik \& Kov\'acs 1998). 
This transformation yields $0.65M_{\odot}$ mean mass of the 40 similar metallicity 
variables with $0.016M_{\odot}$ rms scatter. Thus it is concluded that most 
of the RRab stars belong to a very homogeneous population with mass 
dispersion in accordance with evolutionary model predictions. It can be also 
seen in Fig.~4 that these stars are already in the evolved (redward) phase 
of their horizontal branch evolution. 

The 7 least metal-poor stars comprise a more heterogeneous group, with 
$-1.01\pm0.18$~dex and $0.60\pm0.031M_{\odot}$ mean metallicity and mass, 
respectively.
%
%
\section{Conclusion}
The unusual properties possessed by $\omega$~Centauri already cast some doubt 
about its status as an ordinary globular cluster. Detecting such an anisotropy 
in the spatial distribution of the different metallicity stars as shown in this 
Letter can also hardly be explained in the context of globular cluster evolution 
without any external influence. 

If the metallicity spread had been caused by long lasting continuous or discrete
phases of star-birth periods during the early phase of the cluster's evolution, 
then, in spite of the long relaxation time of the cluster, no initial spatial 
metallicity anisotropy could have been preserved so clearly. It seems that 
there is no mechanism which would date back the origin of both the chemical 
inhomogeneity and its special spatial distribution to the early history of the 
cluster. Some more recent effect is needed in order to explain the phenomenon. 
An explanation may arise from assuming that $\omega$~Centauri is a not-too-old 
merger of two different characteristics stellar clusters. Its Galactic location 
in the vicinity of the boundary between the halo and the disk may have given 
chance for close approach of different population objects in the relatively 
recent past. However, the different kinematic properties of different population 
objects make their merging, even in case of very close approach, unlikely. 
Moreover, in the scope of this picture it also cannot be explained why the 
distribution of the metallicity frequency is not clearly bimodal (see e.g. SK) 
as it would have been expected if two monometallic clusters merge. The fact, 
that not only the more metal-rich stars are located in one side of the cluster
but also the most metal-poor ones in the other, whereas stars not belonging to 
the metallicity extremes are uniformly distributed is also difficult to understand 
in this case. However, we recall that the merging hypothesis was also suggested 
to explain chemical and kinematical distributions and connections by Norris et~al. 
(1996, 1997). 

Alternatively, we should also speculate on that one side of the cluster has been 
polluted by a larger-mass chemically-evolved composition cloud, however, no model 
calculation of such situation can be found in the literature. Thus it is not 
evident whether large enough amount of accretion onto the stellar surfaces should 
then really occur. (In a recent paper Cox (1998) also suggests surface accretion 
as a possible explanation of special period changes observed in cluster variables.) 
This hypothesis would answer why there are no very metal-poor stars in the more 
metal-rich side of the cluster. None of the above explanations can, however, answer 
why the segregation of the metallicity extremes is evident only in the case of the 
bright giants and cannot be detected in the subgiant sample.

Both the ND and SK metallicity values of the giants are affected by evolutionary 
effects since the metallicity calibrations used are not valid for the $\sim$20\% 
evolved, AGB objects. For these stars the given metallicities underestimate the 
true values. Therefore, it is possible that the spatial separation in the 
metallicity distribution also reflects differences in the evolutionary states 
of the stars. But this evolutionary effect does not affect the metal rich sample 
significantly, thus it cannot account for their segregation.

The similar mass and metallicity values found for the bulk of the RR~Lyrae 
stars indicate that the most numerous population of the system resembles 
the ordinary globular clusters well.

Whatever the truth is, $\omega$~Centauri remains undoubtedly the most fascinating 
globular cluster in our Galaxy which is a great challenge for both observations and 
theoretical works in an attempt to reach perfect understanding. 
%
%
\acknowledgements
The supports of OTKA grants T--19640, T--24022, and AKP 96/2--4122,1 are acknowledged.
I wish to thank G\'eza Kov\'acs for providing me the Fourier decompositions of the 
OGLE observations, and B\'ela Szeidl and Lajos Bal\'azs for helpful and interesting 
discussions. I'm also grateful to  L\'aszl\'o Szabados for reading and correcting the
manuscript.
%
%

\newpage
%
%
\centerline{\bf{FIGURE CAPTIONS}}

\figcaption[om1.eps]{The metallicity map of 359 bright ($V\le 12.75$~mag) 
$\omega$ Centauri giants. Filled and open circles are for stars with 
[Fe/H]~$\le-1.75$ and [Fe/H]~$\ge-1.25$, respectively. Dashed line indicates 
the photometric minor axes, which coincides with the rotation axis of the cluster 
within a few degrees. The most and the least metal-poor stars are clearly 
separated along galactic latitudes. The apparent distance between the centers 
of the selected samples is 6.2'. The most metal-rich stars are located in that 
side of the cluster which is towards the galactic plane, whereas the most 
metal-poor ones are on the halo side. Galactic coordinates (J2000) are used.
\label {Fig. 1}}

\figcaption[om2.eps]{Histogram of the distances between the centers ($\Delta C$)
of the extreme metallicity groups obtained from 5000 simulations. In each 
simulation the observed metallicities of 464 bright giants ($V\le 13$~mag) were 
randomly distributed among their positions. The arrow indicates the observed  
$\Delta C$ value. It has only 0.1\% probability to occur such a displacement 
as observed.
\label {Fig. 2}}

\figcaption[om3.eps]{The relative galactic latitude ($\Delta {\rm b}$) of the 
stars measured from the position of the cluster center are plotted against 
magnitude. As in Fig. 1, dots and circles indicate the most and least metal-poor 
stars, respectively. It can be also seen here that in the case of the bright 
giants the separation of the two samples along galactic latitude is clearly 
evident, whereas the fainter samples show no sign of similar metallicity 
segregation.
\label {Fig. 3}}

\figcaption[om4.eps]{HR diagram of the $\omega$ Centauri RRab stars. Filled 
circles denote variables with $-1.30 \geq {\rm [Fe/H]} \geq-1.70$. Open circles 
show the positions of the more and the less metal-poor stars, the circle size 
is proportional to the [Fe/H] value. For comparison, a segment of Dorman's (1992) 
horizontal branch evolutionary track ([Fe/H]$=-1.48$, $M=0.66M_{\odot}$) is also 
drawn by dashed line.
\label {Fig. 4}}

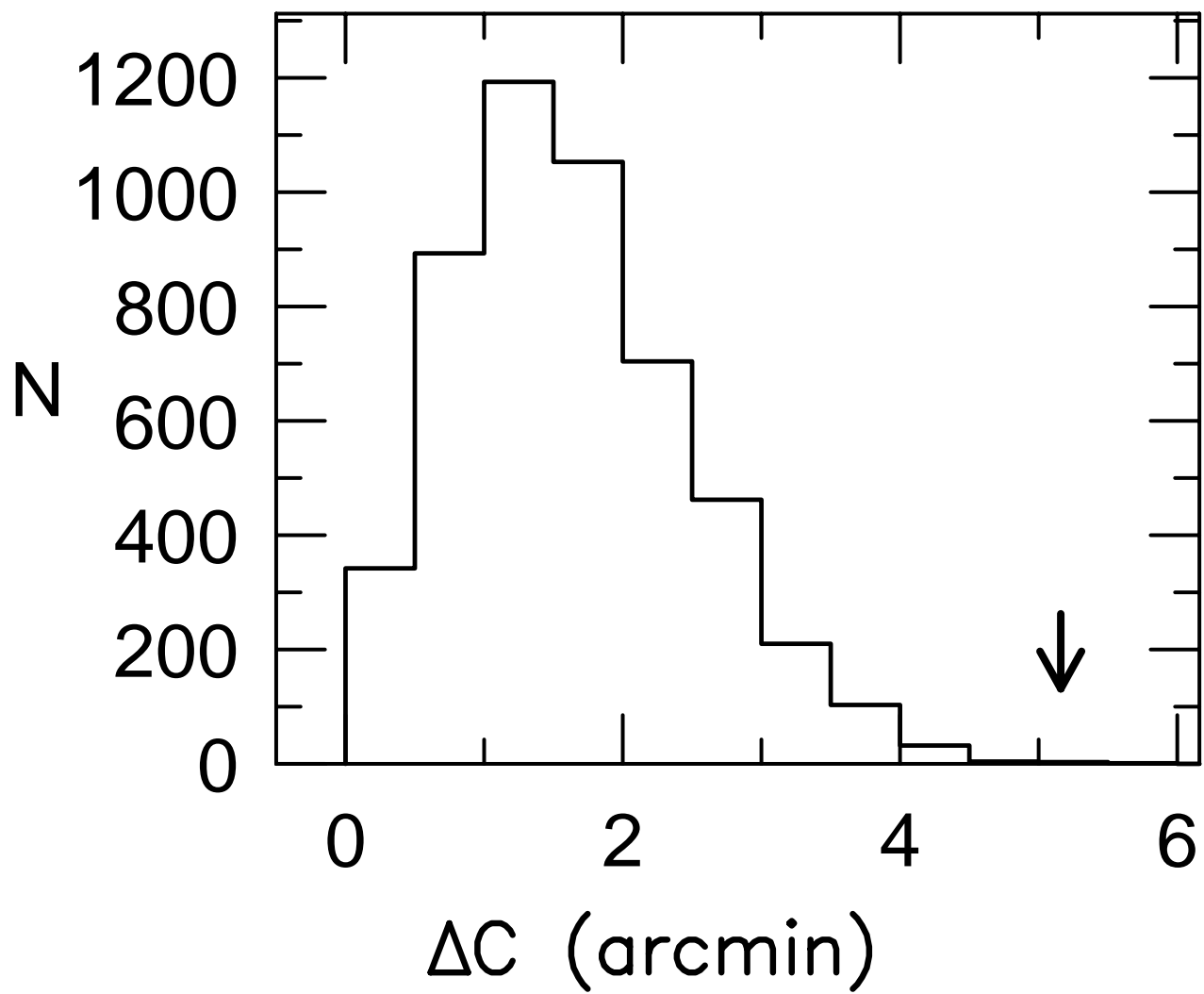
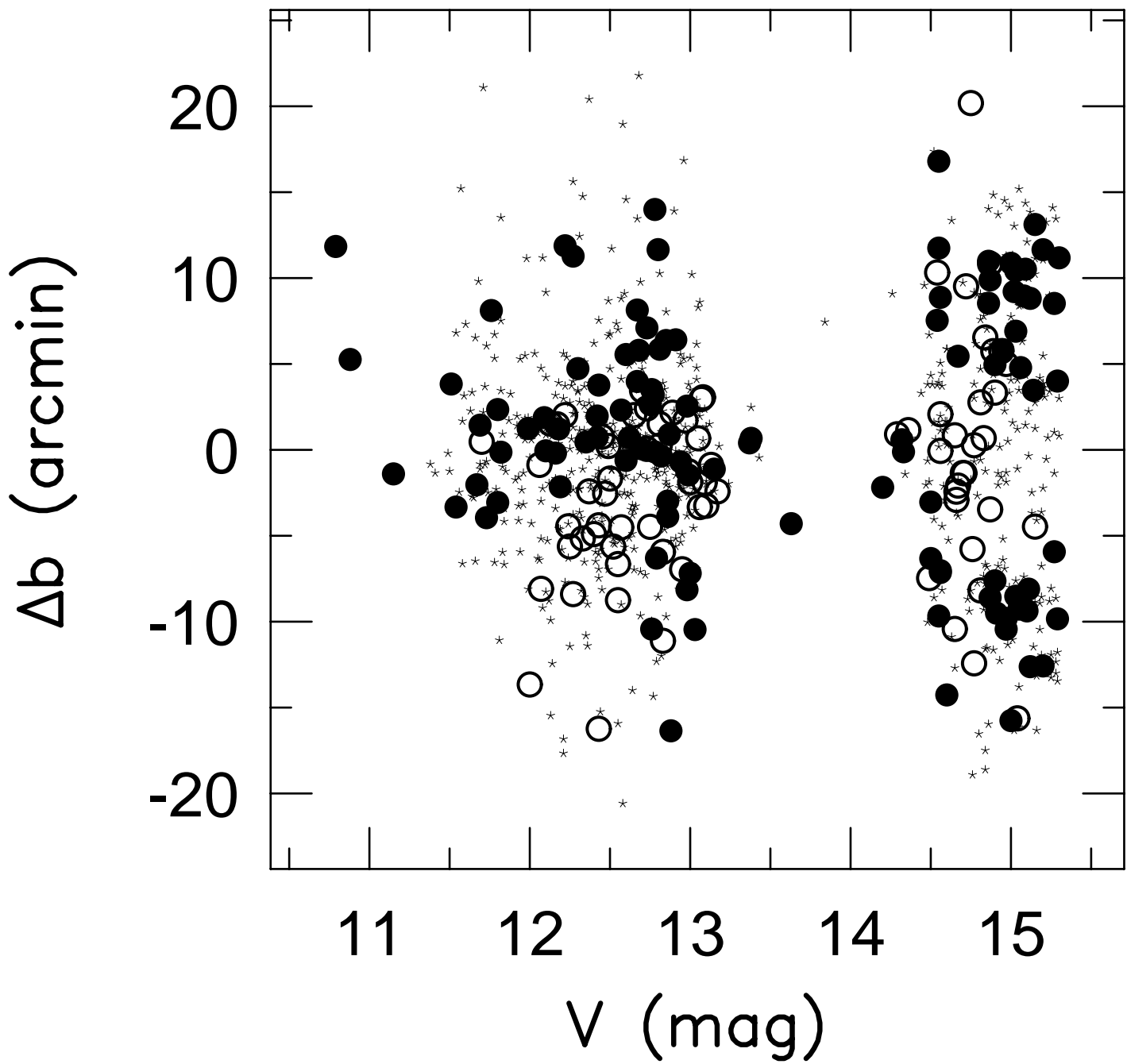
\end{document}